\documentclass[onecolumn]{IEEEtran}
\usepackage{ifpdf}
\usepackage{cite}
\ifCLASSINFOpdf
  \usepackage[pdftex]{graphicx}
\else
  \usepackage[dvips]{graphicx}
\fi
\usepackage{amsmath}
\usepackage{algorithmic}
\usepackage{array}
\usepackage{fixltx2e}
\usepackage{stfloats}
\begin{document}

\title{A Soft SIMD Based Energy Efficient\\Computing Microarchitecture}
\author{
	\IEEEauthorblockN{
		Pengbo Yu\IEEEauthorrefmark{1}, 
		Alexandre Levisse\IEEEauthorrefmark{1}, 
		Giovanni Ansaloni\IEEEauthorrefmark{1}, 
		David Atienza\IEEEauthorrefmark{1} \\
		Mohit Gupta\IEEEauthorrefmark{2} 
		Evenblij Timon\IEEEauthorrefmark{2} 
		Francky Catthoor\IEEEauthorrefmark{2}
		}
		
	\IEEEauthorblockA{\IEEEauthorrefmark{1}Embedded Systems Laboratory (ESL), EPFL, Lausanne, Switzerland\\ Email: {\{pengbo.yu, alexandre.levisse, giovanni.ansaloni, david.atienza\}}@epfl.ch}%

	\IEEEauthorblockA{\IEEEauthorrefmark{2}Interuniversity Microelectronics Centre (IMEC), Leuven, Belgium\\ Email: {\{Mohit Gupta, Evenblij Timon, Francky Catthoor\}}@imec.be}
       }%

\maketitle
\begin{abstract}
The ever-increasing size and computational complexity of today's machine-learning algorithms pose an increasing strain on the underlying hardware. In this light, novel and dedicated architectural solutions are required to optimize energy efficiency by leveraging opportunities (such as intrinsic parallelism and robustness to quantization errors) exposed by algorithms. We herein address this challenge by introducing a flexible two-stages computing pipeline. The pipeline can support fine-grained operand quantization through software-supported Single Instruction Multiple Data (SIMD) operations. Moreover, it can efficiently execute sequential multiplications over SIMD sub-words thanks to zero-skipping and Canonical Signed Digit (CSD) coding. Finally, a lightweight repacking unit allows changing the bitwidth of sub-words at run-time dynamically. These features are implemented within a tight energy and area budget. Indeed, experimental results showcase that our approach greatly outperforms traditional hardware SIMD ones both in terms of area and energy requirements. In particular, our pipeline occupies up to 53.1\% smaller than a hardware SIMD one supporting the same sub-word widths, while performing multiplication up to 88.8\% more efficiently.
\end{abstract}

\begin{IEEEkeywords}
Hardware Software Co-design, Energy Efficient Computing, Single Instruction Multiple Data, Edge Machine Learning. 
\end{IEEEkeywords}
\IEEEpeerreviewmaketitle

\section{Introduction}
\label{sec:introduction}
Machine Learning (ML) has fostered a revolution in computing, impacting a plethora of domains ranging from healthcare \cite{qayyum2020secure} to finance \cite{ozbayoglu2020deep}. Nonetheless, the high computational requirements of ML applications pose a challenge to their deployment, especially when targeting resource-constrained devices \cite{murshed2021machine}.

To address the deployment challenges mentioned above, two main optimization avenues have been proposed. They take advantage of a) the high degree of parallelism offered by ML algorithms and b) their robustness towards low-range data representation. Our research contribution also leverages these opportunities, declining them towards the design of a dedicated computing pipeline. The pipeline is optimized for parallel, small-bitwidth arithmetics, which supports flexible Single Instruction Multiple Data (SIMD) formats. It efficiently implements the software SIMD (Soft SIMD \cite{kraemer2007softsimd}\cite{catthoor2010ultra}) computing paradigm and a Canonical Signed Digit (CSD \cite{oudjida2014binary} \cite{nigam2022hardware}) representation of operands. 

Our solution features a fine-grained configurability of the adopted data bitwidths, which can be defined at run-time according to the desired computation precision, e.g., adapting to the robustness of different layers in an ML network \cite{ponzina2021flexible}\cite{young2021transform}. It can perform the parallel multiplication of a multiplier value with several multiplicands, the most prominent operation in ML applications. Moreover, it allows seamless transitions among different SIMD formats using a dedicated data packing stage. Our design is extremely parsimonious in terms of area and energy resources, paving the way for its integration as a near-memory accelerator interfacing memory banks \cite{sudarshan2022critical}, hence harnessing the regularity of computations that characterize ML applications. 

In summary, our contributions are as follows:
\begin{itemize}
\item{} We introduce a novel two-stage pipeline supporting the Soft SIMD and CSD coding paradigms to support parallel multiplications, as well as data-repacking to bridge between SIMD formats.
\item{} Through detailed post-synthesis analyses, we show that our approach offers superior flexibility and area/energy efficiency with respect to traditional solutions based on hardware SIMD (Hard SIMD). In particular, our design requires up to 53.1$\%$ less area than an equivalent Hard SIMD implementation. Moreover, our design consumes up to 88.8$\%$ less energy to perform a multiplication.
\end{itemize}

The paper proceeds as follows: in Section \ref{sec:background}, we introduce the foundation notions providing the rationale for our design choice. Then, the proposed pipeline is detailed in Section \ref{sec:design}. Comparative experimental evaluations are provided in Section \ref{sec:experiments}, while Section \ref{sec:conclusion} concludes the paper.

\section{Background}
\label{sec:background}

\subsection{Software SIMD}

Single Instruction Multiple Data is a well-known approach to enhance parallelism, hence performance, when executing regular computation patterns. Indeed, mainstream Instruction Set Architectures (ISAs) do define SIMD extensions, e.g., in the form of ARM Neon  \cite{reddy2008neon}, or x86 AVX \cite{kusswurm2014advanced} instructions. Their hardware implementation entails using wide registers hosting a fixed number of sub-words (of fixed bitwidth) and the corresponding parallel functional units performing vector operations among registers. 

\begin{figure}[htbb]
  \vspace{-0.0cm}
  \centering
  \includegraphics[width = .5\columnwidth{}]{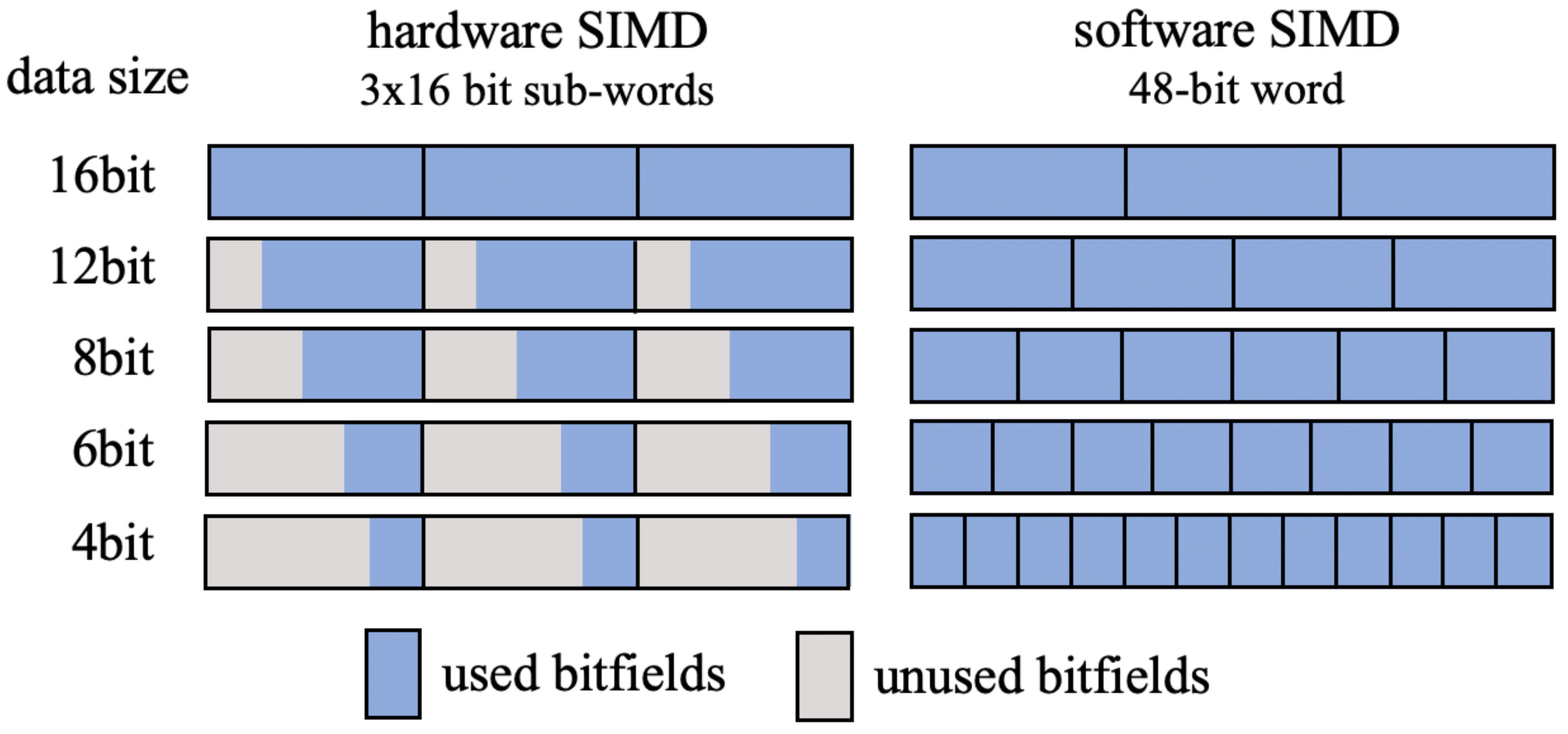}
  \vspace{-0.1cm}
  \caption{Soft SIMD can adapt the size of sub-words bitwidths to data size requirements, maximising utilization and word-level parallelism.}
  \label{fig:01}
  \vspace{-0.0cm}
\end{figure}

Soft SIMD, proposed in \cite{kraemer2007softsimd}\cite{catthoor2010ultra}, challenges this paradigm, allowing arbitrary partitioning of vector registers in sub-words, as illustrated in Figure \ref{fig:01}. Soft SIMD formats can then be configured in software at run-time (hence the name), ensuring that operations on a sub-word would not cause side effects on other sub-words, even in the case of positive/negative overflows. Soft SIMD is particularly appealing when a fine-grained control of bitwidths is beneficial, employed bitwidths are small, and/or the desired data representation changes in different computation phases. These features truly characterize ML applications, as discussed in Section \ref{sec:introduction}.

Soft SIMD can be embodied along two alternative approaches: reserving bit locations in-between sub-words (guardbits) or adopting configurable carry generation at sub-word boundaries in arithmetic units with low additional hardware cost and timing path increase \cite{kraemer2007softsimd}\cite{catthoor2010ultra}\cite{psychou2012sub}. In the implementation presented in this work, we employ a configurable carry generation, but our strategy is also compatible with the approaches using guardbits. 

\subsection{CSD Coding}
Our proposed pipeline performs parallel multiplications as a sequence of arithmetic shifts and additions between multiplicands and accumulators, processing one bit of a multiplier operand each time. As detailed in Section \ref{subsec:mult}, only the shift operation is performed when the multiplier bit is `0'. Therefore, multiple shifts can be coalesced in the same clock cycle when processing bit patterns with trailing zeros such as ``10", ``100", etc. To maximize the ensuing performance benefit, we herein adopt a Canonical Signed Digit (CSD) representation for multiplier values  \cite{oudjida2014binary}\cite{nigam2022hardware}. Such encoding employs three symbols for each digit: `1', `0', and `-', where the latter indicates that the number should be interpreted as negative. As an example, the number ``0-01" in CSD notations equals to $(-4) + 1 = -3$.
In CSD numbers, $\sim{}(\tfrac{2}{3})$ of the digits are zeroes, increasing opportunities for  coalescing multiple shifts.

\section{Soft SIMD Microarchitecture}
\label{sec:design}

\subsection{Soft SIMD Arithmetic Pipeline}\label{subsec:AA}

\begin{figure}[htbp]
  \vspace{-0.2cm}
  \includegraphics[width=0.5
  \columnwidth{}]{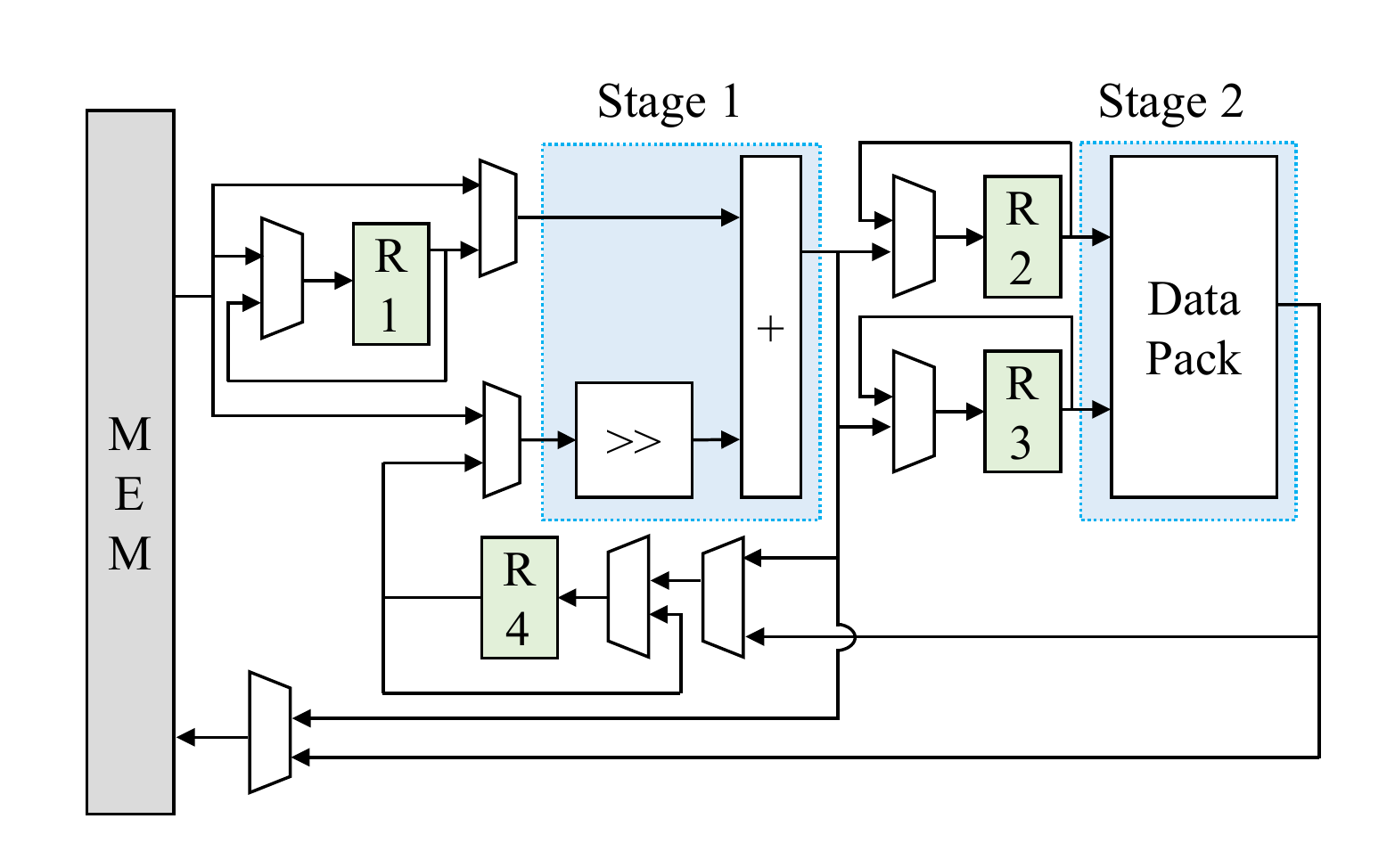}
  \centering
  \vspace{-0.2cm}
  \caption{Block scheme of the 2-stages Soft SIMD pipeline. Stage 1 is devoted to arithmetic operations, Stage 2 to data repacking.}
  \label{fig:02}
\end{figure}

A block scheme of the hardware design is illustrated in Figure \ref{fig:02}. It is composed of two pipeline stages. The first one is dedicated to parallel shift-add operations, hence sequentially performing multiplications. As we will show in Section \ref{sec:experiments}, such an approach leads to higher efficiency (especially for small-bitwidth operands) with respect to using combinatorial multipliers because it requires simpler (leaner, faster) logic. The second pipeline stage is instead devoted to data packing, namely, to the translation of values between bitwidths across computation phases adopting different SIMD formats. This second stage can be bypassed if no data conversion is required. Details on datapath stages are provided in the following sub-sections.

\subsection{Soft SIMD Multiplication}
\label{subsec:mult}
The first pipeline stage allows the multiplication of values represented in fixed-point $Q1.X$ notation, i.e., with one leading bit for the integer parts and the rest devoted to the fractional part. Operations are performed parallel among a multiplier, encoded in CSD notation, and several multiplicands are represented in 2’s complement notation. 

\begin{figure}[t]
  \vspace{-0.0cm}
  \centering
  \includegraphics[width=.4\columnwidth{}]{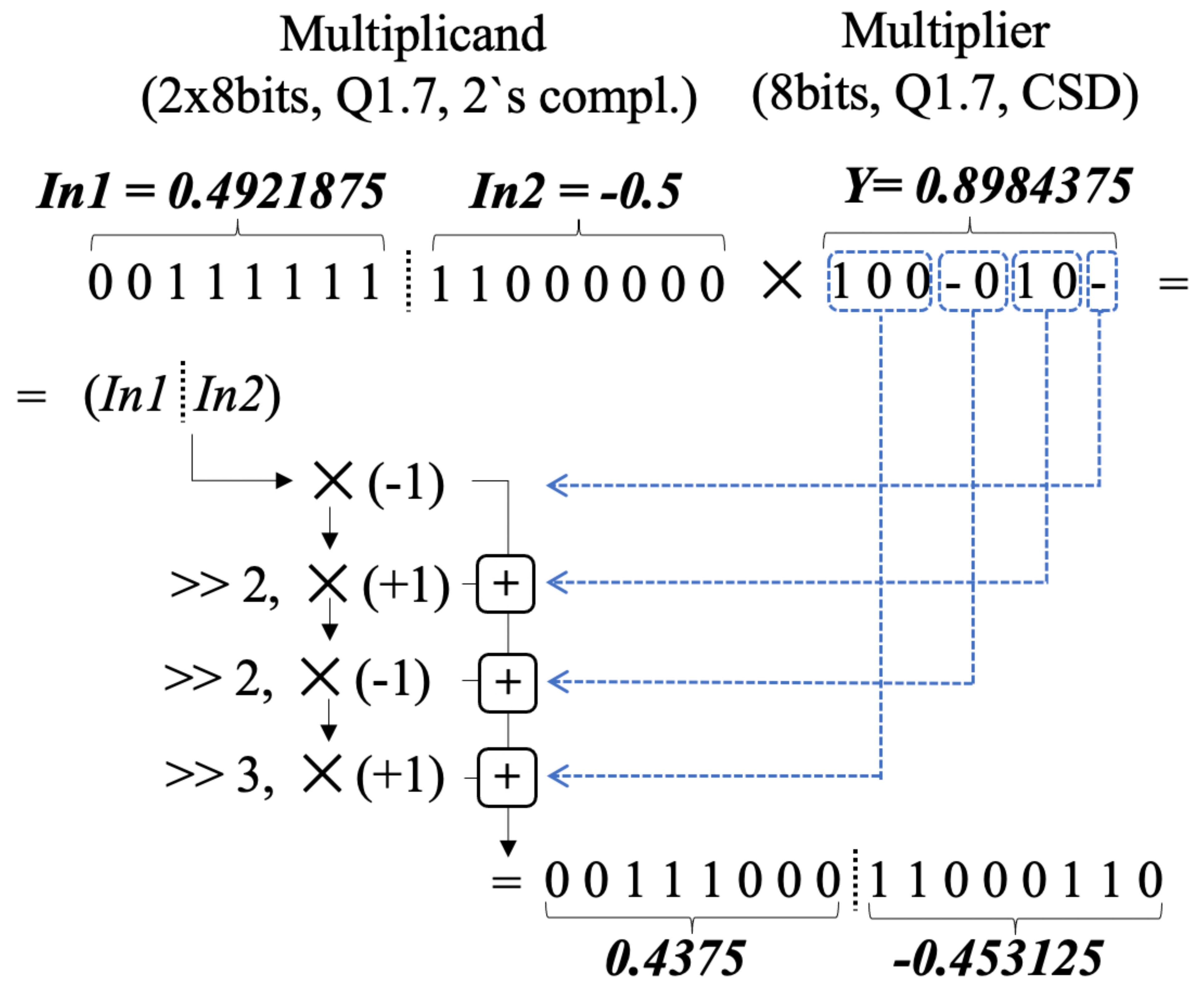}
  \vspace{-0.2cm}
  \caption{Multiplication steps. Operands are a SIMD multiplicand Q1.X 2's complement format (x=7 in the example), and a multiplier in Q1.Y CSD format (y=7 in the example, 01110011 before CSD). At each step, partial results are computed by right shift and add operations. Multiple bit-fields of the multipliers containing trailing zeros can be processed at one. In the example, only three additions are required.   }
  \label{fig:03}
  \vspace{-0.5cm}
\end{figure}

The employed algorithm is shown in Figure \ref{fig:03} using an example of a $Q1.7$ multiplier and two 8-bit multiplicands stored as Soft SIMD sub-words, themselves represented as $Q1.7$ values. Notice that since the bitwidth of results and that of the multiplicands are the same, truncation errors can and do occur. Nonetheless, these are negligible even for very constrained bitwidths, e.g., approximately $1\%$ in the shown 8-bit example. 

\begin{figure}[t]
  \vspace{-0.3cm}
   \centering
  \includegraphics[width=0.5\columnwidth]{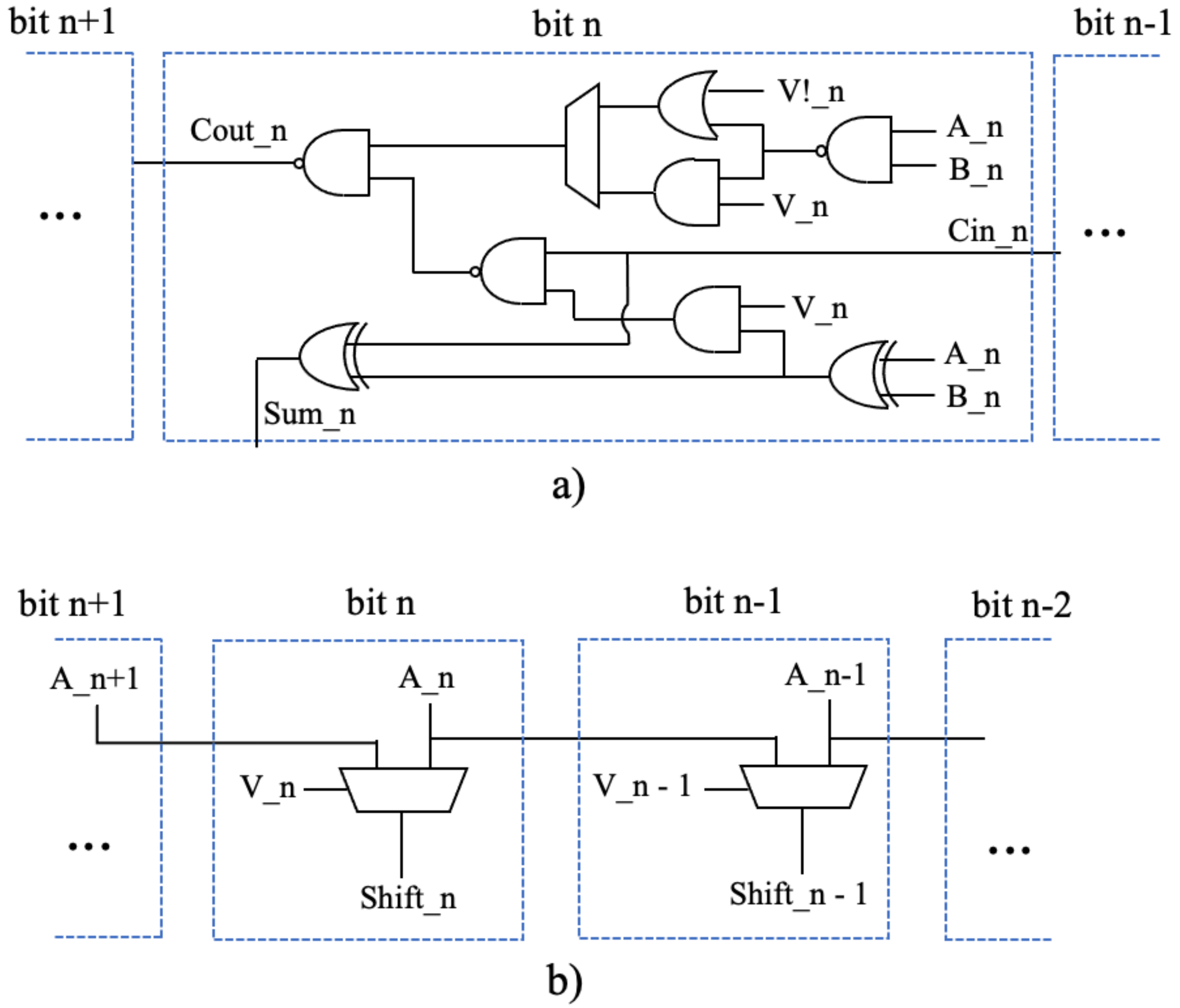}
  \vspace{-0.5cm}
  \caption{(a) 1-bit slice of the configurable adder circuit (b) structure of the configurable shifter. In both cases, V\_x equals `0' in bit positions corresponding to the MSB of sub-words, and `1' otherwise.}
  \label{fig:04}
  \vspace{-0.4cm}
\end{figure}

Figure~\ref{fig:03} showcases that right shifts, complements (multiplications by $-1$), and additions are required to implement multiplication. Such operations must not result in interference from one sub-word to the next. To this end, sub-word boundaries prevent overflow from passing in additions and provide +1 for the next sub-word in subtractions in our design by expandable control signals. The employed circuit is shown in Figure \ref{fig:04}-a. Since the carry logic operation of the multi-bit adder is performed simultaneously, the increase in the time path is very small. As the shift operation, the Most Significant Bit (MSB) of sub-words propagates the corresponding bit position of the input, hence implementing sign extension (Figure \ref{fig:04}-b). It has to be noted that muxes can be employed selectively, depending on the supported bitwidths. Indeed, no mux is required if a bit position is never the MSB of a sub-word for all supported Soft SIMD formats. Such a strategy can scale to support multiple shifts by employing further combinatorial stages of 1-bit muxes. In this way, multi-bit patterns with trailing zeros can be processed in a single clock cycle, as shown in Figure \ref{fig:03}. In our design, we support up to 3-bit patterns, as more extensive sequences of consecutive zeros are rare and do not justify the additional logic.

\subsection{Data Packing Unit}
\begin{figure}[t]
  \vspace{-0.1cm}
  \centering
  \includegraphics[width = .3\columnwidth{}]{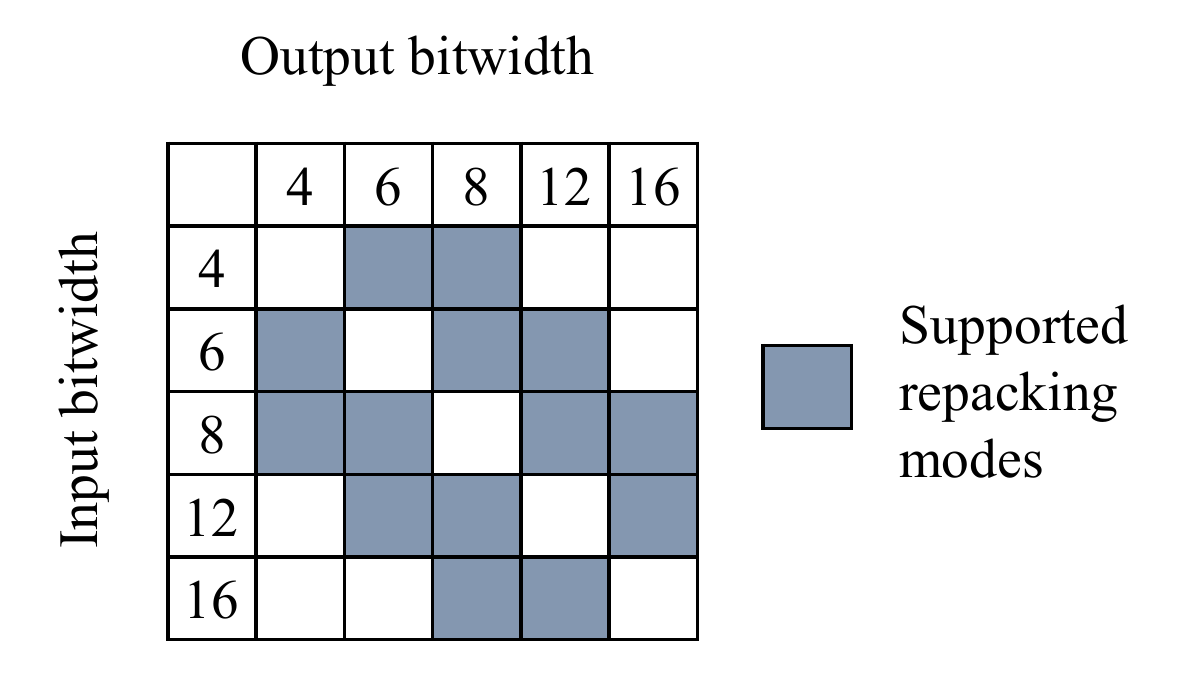}
  \vspace{-0.3cm}
  \caption{SIMD format conversions supported in our data pack design.}
  \label{fig:05}
  \vspace{-0.2cm}
\end{figure}

The role of the data packing unit is to bridge across SIMD formats. To this end, a crossbar \cite{ raghavan2007customized}  is employed to connect bits in different bit ranges of the Stage2 inputs (registers R2, R3 in Figure \ref{fig:02}) to the Stage2 output (registers R4 or write back to memory). The hardware resources required by the crossbar depend on the width of the datapath and the set of employed Soft SIMD formats. In the design investigated in Section \ref{sec:experiments}, we considered a 48-bits datapath and sub-words of 4, 6, 8, 12, and 16 bits. We support many conversions between modes, as indicated in Figure \ref{fig:05}. Finally, the entire stage can be bypassed if no change in sub-word format is required.

\section{Experiment and Result}
\label{sec:experiments}

\subsection{Experiment Setup}
Across experiments, we target a 48-bit pipeline. To investigate its performance, we compare its efficiency with that of Hard SIMD solutions employing combinatorial multipliers. The first baseline Hard SIMD architecture supports sub-word configurations 4, 6, 8, 12, and 16 bits, while the other only supports 8 and 16 bits. In all cases, timing, area, and energy characterizations are performed on post-synthesis data based on a 28nm technology library under varying timing constraints. 

\subsection{Area Evaluation}

\begin{figure}[b]
\vspace{-0.5cm}
  \includegraphics[scale=0.5]{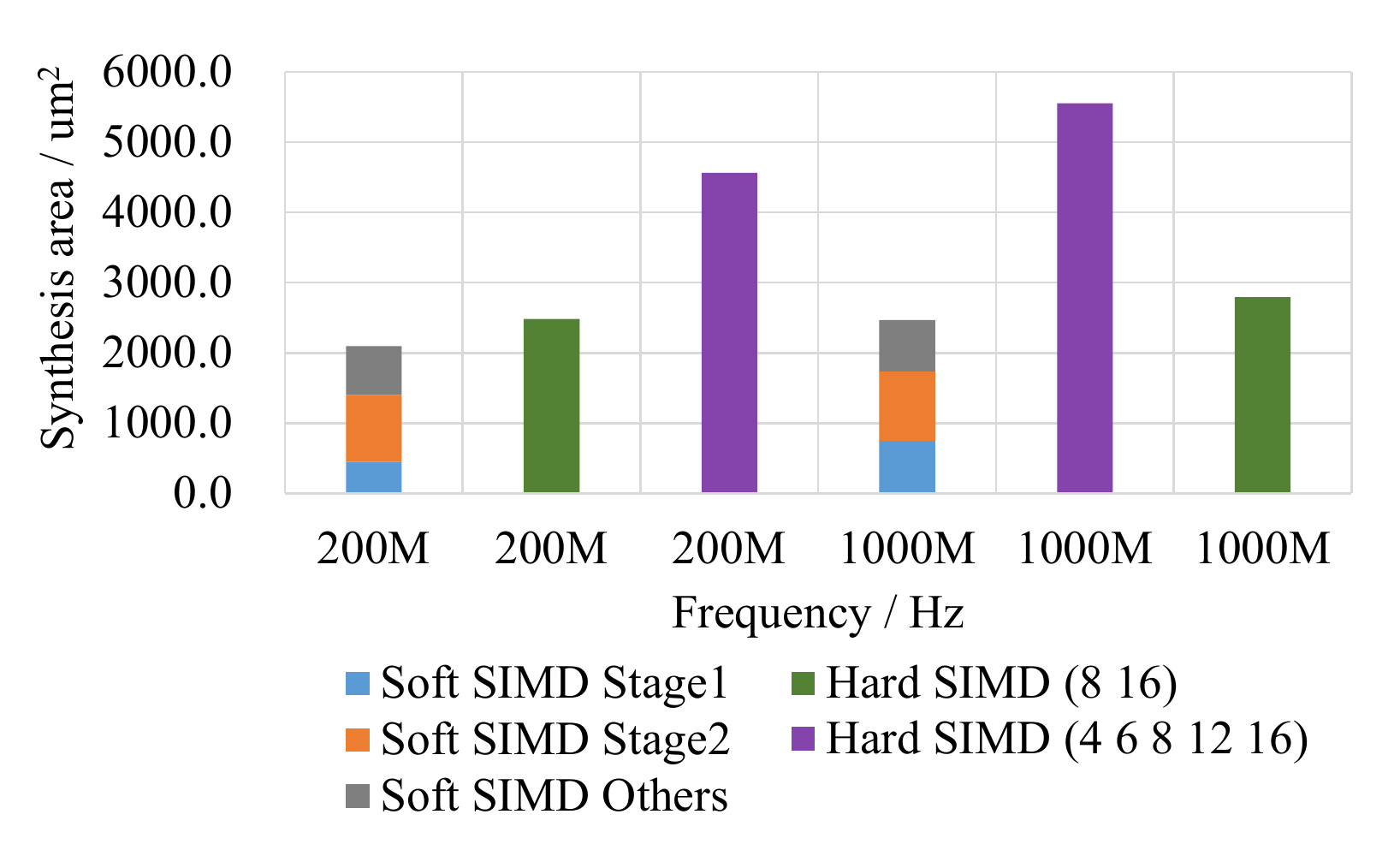}
  \centering
  \caption{Area of Soft SIMD and Hard SIMD pipelines, when synthesized with either a 200MHz or 1GHz timing constraint.}
  \label{fig:06}
\end{figure}

\begin{figure}[htbp]
\vspace{-0.1cm}
  \includegraphics[scale=0.6]{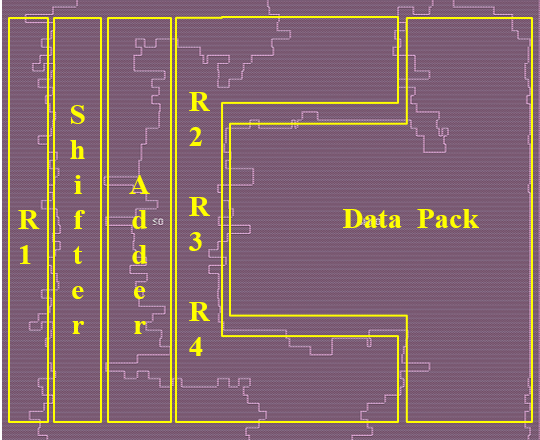}
  \centering
  \caption{Design layout after place-and-route.}
  \label{fig:07}
  \vspace{-0.2cm}
\end{figure}

\begin{figure}[htbp]
\vspace{-0.2cm}
  \includegraphics[scale=0.5]{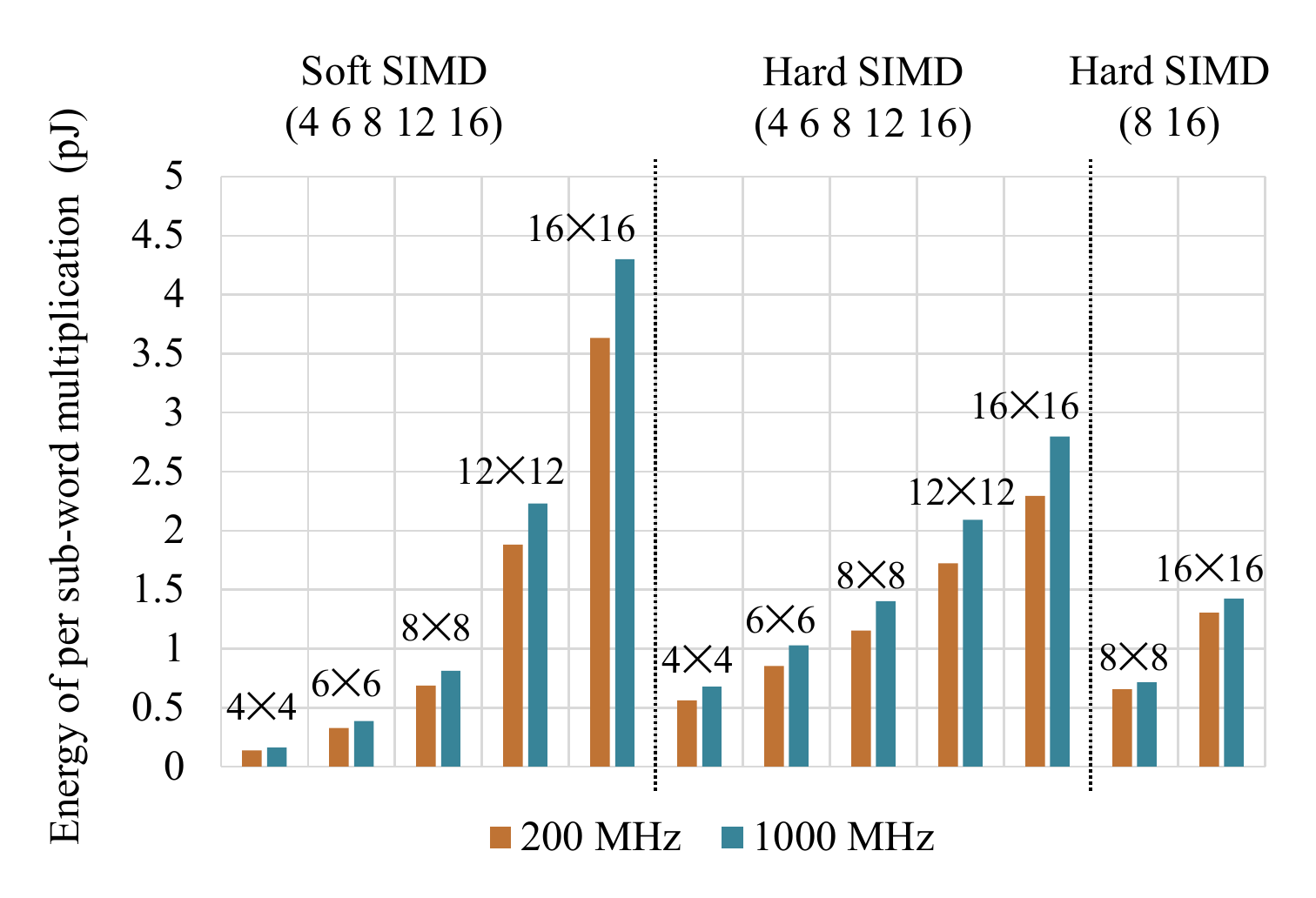}
  \centering
  \caption{Energy requirement for 1 sub-word multiplication, for selected Soft SIMD and Hard SIMD configurations and different synthesis timing constraints.}
  \label{fig:08}
\end{figure}

\begin{figure}[htbp]
\vspace{-0.0cm}
  \includegraphics[scale=0.4]{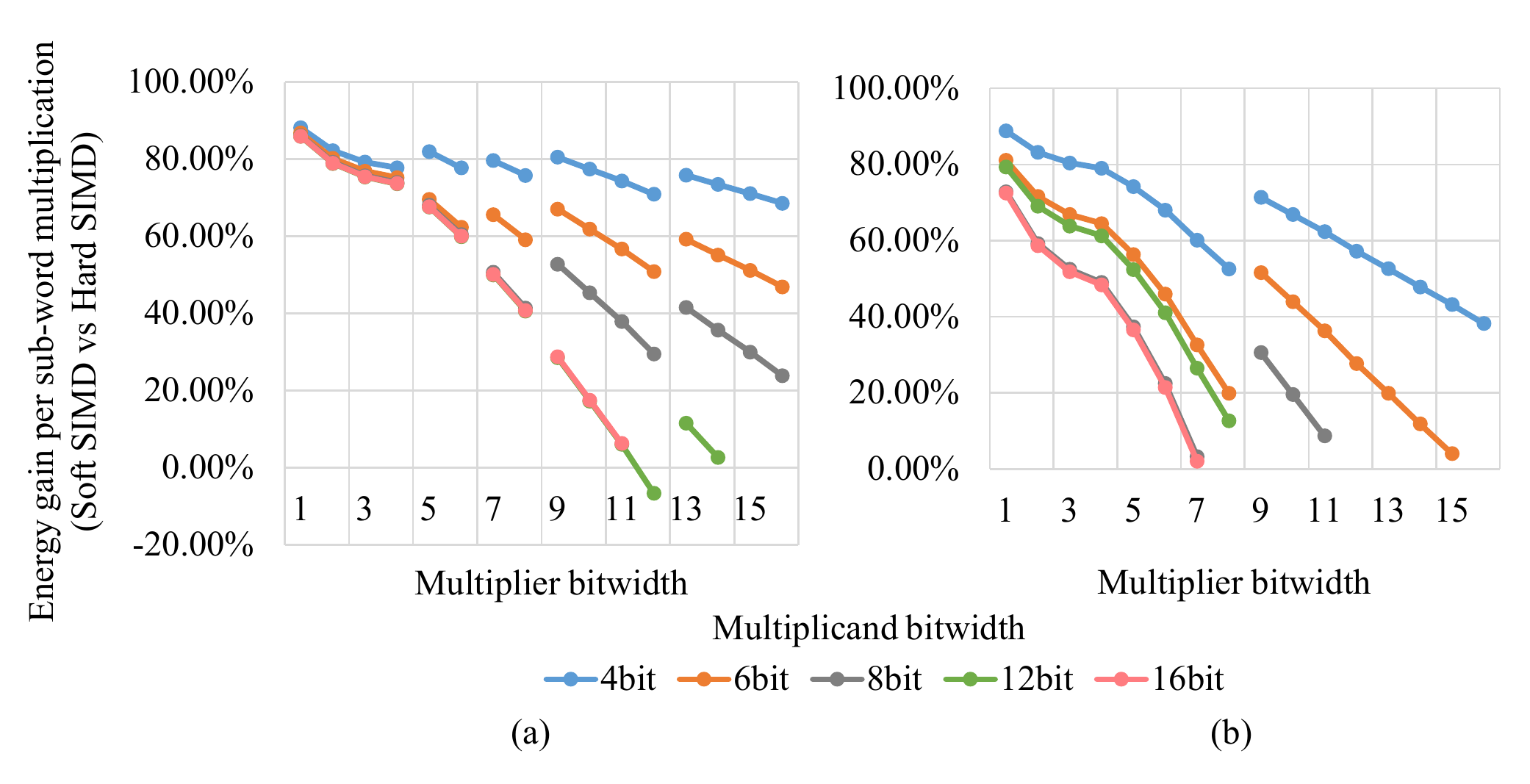}
  \centering
  \vspace{-0.2cm}
  \caption{(a) Energy gain of Soft SIMD with respect to (a) Hard SIMD (4 6 8 12 16) and (b) Hard SIMD (8 16), varying multiplicand and multiplier bitwidths, at 1000MHz.}
  \label{fig:09}
  \vspace{-0.0cm}
\end{figure}

\begin{figure}[htbp]
\vspace{-0.0cm}
  \includegraphics[scale=0.4]{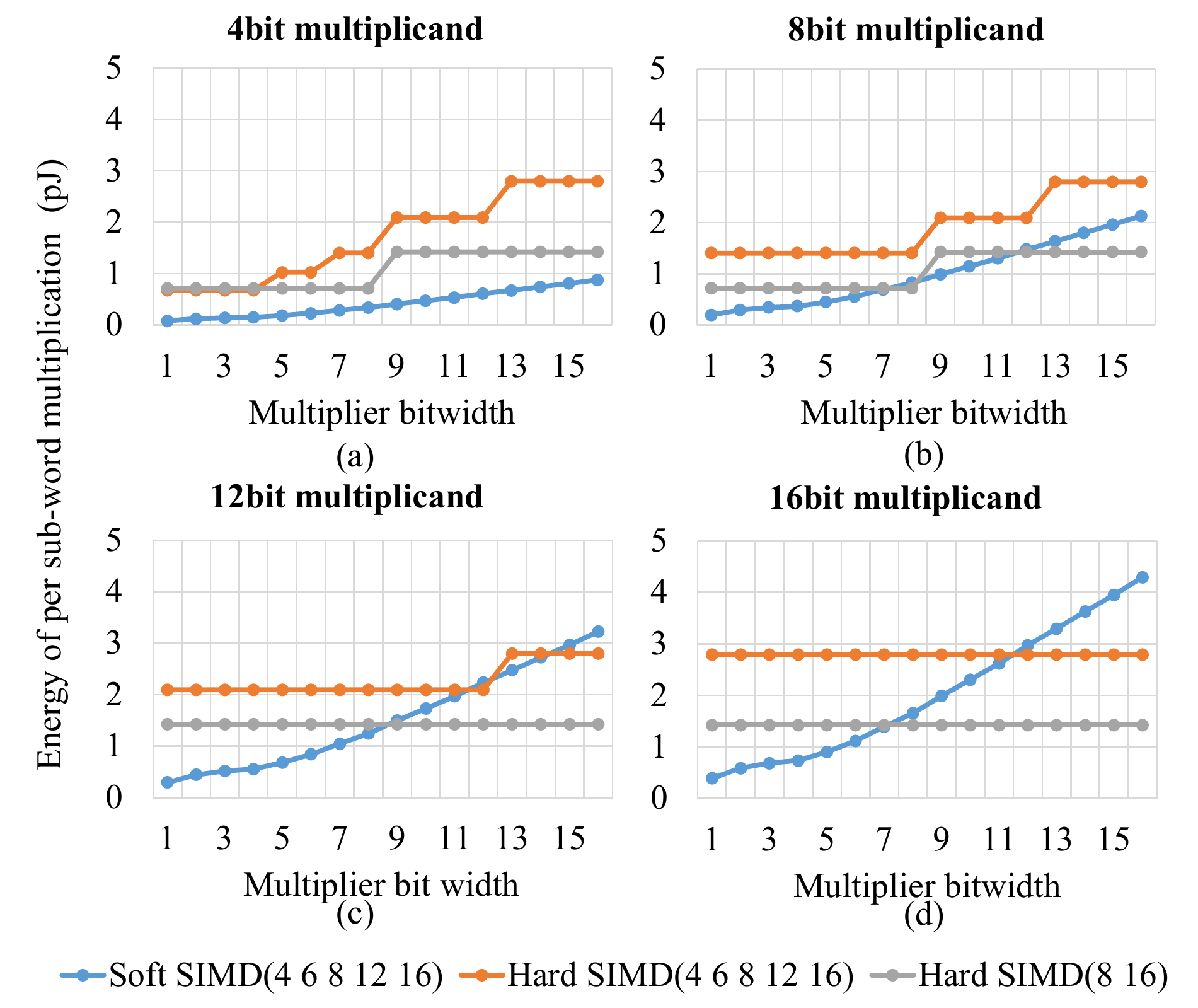}
  \centering
  \caption{Average energy of per sub-word Multiplication across different scenarios at 1000MHz.}
  \label{fig:10}
  \vspace{-0.3cm}
\end{figure}

Figure \ref{fig:06} compares the area of Hard SIMD and Soft SIMD solutions. As expected, as Soft SIMD does not require a resource-hungry combinatorial multiplier, its area requirements are markedly lower with respect to Hard SIMD for the same supported bitwidths (less than half across different timing constraints). Moreover, the area efficiency of Soft SIMD is not matched by Hard SIMD even when only 8- and 16-bit sub-words are supported in the latter, which still remains more than 10\% larger in all cases. In addition, stage 2 of Soft SIMD
remains basically constant at different frequencies while stage 1 and other parts (registers, etc.) grow with frequency. The design layout is presented in Figure \ref{fig:07}.

We evaluated the efficiency of the proposed pipeline in terms of the energy (pico-Joules) required to perform a multiplication among multiplicands and multipliers of varying bitwidths. In all cases, the bitwidth of the result matches that of the multiplicand. Absolute energy results for selected configurations are shown in Figure \ref{fig:08}, highlighting that Soft SIMD achieves better energy efficiency for widths smaller than 8 bits. Moreover, Soft SIMD can seamlessly support a large number of data widths, while in the Hard SIMD case, flexibility comes at a hefty cost in terms of efficiency, as the comparison between the two Hard SIMD solutions in the $8\times{}8$ and $16\times{}16$ configurations highlights.

A comprehensive view of the efficiency of our proposed pipeline and that of the compared baseline designs is presented in Figure \ref{fig:09}. This figure plots the energy gain of Soft SIMD with respect to the two Hard SIMD implementations when executing multiplications of different multiplicands and multipliers widths. These results show that import gains are obtained, especially for small bitwidths. Moreover, in these comparisons, the plotted series present discontinuities when the multiplicand width exceeds the size of the Hard SIMD sub-words, i.e., between 8 and 9 bits in Figure \ref{fig:09}, on the left side.

Finally, as Figure \ref{fig:10} shows, the benefits of Soft SIMD derive from its graceful scaling to different bitwidths in terms of energy requirements. Conversely, even though Hard SIMD often enables the support of many SIMD formats, this flexibility does not lead to efficiency increases. In fact, the Hard SIMD (4 6 8 12 16) solution consistently underperformed the simpler Hard SIMD (8 16) option in our experiments.

In summary, Soft SIMD achieves significant energy gains (up to 88.8\%) for multiplications with small bit-width operands compared to traditional Hard SIMD alternatives. Such characteristics are of great interest for algorithms amenable to aggressive quantization, such the machine learning ones.

\section{Conclusion}
\label{sec:conclusion}
\subsection{Energy Efficiency Evaluation}

This paper presents a novel computing pipeline, leveraging the Soft SIMD paradigm and CSD encoding to achieve ultra-high energy efficiency when performing arithmetic operations among aggressively quantized values. The pipeline features the skipping of trailing '0' digits in multiplier operands, seamless support of a wide range of quantization levels, and a repacking unit to bridge across SIMD formats.

Experiments showcase that our Soft SIMD design greatly outperforms Hard SIMD alternatives when executing multiplications among small-bitwidth operands, highlighting energy gains of up to 88.8\% in energy, while requiring fewer hardware resources.
\vspace{0.3cm}
\bibliographystyle{IEEEtran}
\bibliography{reference.bib}
\end{document}